\documentclass
[superscriptaddress,secnumarabic,amssymb,amsmath,nobibnotes,aps,prd,showkeys,showpacs,twocolumn,nofootinbib]{revtex4}%
\usepackage{graphicx}
\usepackage{bm}
\usepackage{amsmath}
\usepackage{amsfonts}
\usepackage{amssymb}
\usepackage{epstopdf}%
\providecommand{\U}[1]{\protect\rule{.1in}{.1in}}
\newcommand{\be}{\begin{equation}}
\newcommand{\ee}{\end{equation}}

\newcommand{\mincir}{\raise
-3.truept\hbox{\rlap{\hbox{$\sim$}}\raise4.truept\hbox{$<$}\ }}
\newcommand{\magcir}{\raise
-3.truept\hbox{\rlap{\hbox{$\sim$}}\raise4.truept\hbox{$>$}\ }}

\begin{document}
\title{On the existence of static spherically-symmetric objects in action-dependent Lagrangian theories}

\author{Julio C. Fabris}
\email{julio.fabris@cosmo-ufes.org}
\affiliation{N{\'u}cleo Cosmo-ufes \& PPGCosmo, Universidade Federal do Esp{\'i}rito Santo,
29075-910, Vit{\'o}ria, ES, Brazil}
\affiliation{National  Research  Nuclear  University  MEPhI,  Kashirskoe  sh.   31,  Moscow  115409, Russia}

\author{Hermano Velten}
\email{hermano.velten@ufop.edu.br}
\affiliation{Departamento de F{\'i}sica, Universidade Federal de Ouro Preto (UFOP), 35400-000, Ouro Preto, MG, Brazil}

\author{Aneta  Wojnar}
\email{aneta.wojnar@cosmo-ufes.org}
\affiliation{N{\'u}cleo Cosmo-ufes \& PPGCosmo, Universidade Federal do Esp{\'i}rito Santo,
29075-910, Vit{\'o}ria, ES, Brazil}


\date{\today}

\begin{abstract}

We study static symmetric solutions in the context of a gravitational theory based on a action-dependent Lagrangian. Such theory has
been designed as a setup to implement dissipative effects into the traditional principle of least action. Dissipation appears therefore 
from the first principles and has a purely geometric origin. An interesting feature of this theory is the existence of a coupling 
four-vector $\lambda_{\mu}$, which in an expanding background is related to cosmological bulk viscosity. General Relativity is recovered with a vanishing $\lambda_{\mu}$. We analyse the existence of equilibrium solutions of static configurations aiming to describe astrophysical objects. We find out that the existence of static spherically symmetric configurations occurs only in the particular scenario with vanishing $\lambda_t$, $\lambda_r$ and $\lambda_{\phi}$ components i.e, $\lambda_{\mu}=\{0,0,\lambda_{\theta},0\}$. Thus, the component $\lambda_{\theta}$ is the unique available parameter of the theory in the astrophysical context. This result severely constrains the existence of this sort of gravitational theories. We proceed then verifying the impact of $\lambda_{\theta}$ on the stability and the mass-radius configurations for a reasonable equation of state for the cold dense matter inside compact stars. We further investigate the relativistic spherical collapse in order to track the structure of geometrical singularities appearing in the theory.

\end{abstract}

\maketitle

\section{Introduction}

The General Relativity (GR) based description of astrophysical and cosmological observables has led to the concept of dark matter and dark energy. While the former
has been evoked to deal with the dynamics at galactic/clusters scales, the latter is assumed to drive the accelerated background expansion of 
Friedmann-Lemaitre-Robertson-Walker (FLRW) universe. 

In order to circumvent the dark matter and dark energy phenomena the idea to extend the description of gravitational interaction beyond GR has set a well-established 
route of investigation in the literature. This strategy is based mainly on the concept of giving up the typical GR gravitational 
Lagrangian density $\mathcal{L}_{g} = R$, where $R$ is the Ricci scalar, to construct the more generic Lagrangian density with the 
form $\mathcal{L}_g\equiv \mathcal{L}(R, {\rm other \,\,terms})$.

Recently \cite{lazo}, within the spirit of searching for new viable geometric structures to describe gravity, {\it Lazo et al} have proposed a covariant generalization of the Herglotz problem. The latter consists in generalizing the minimum action principle via the introduction of action-dependent Lagrangians $S = \int \mathcal{L}(x, \dot{x},S) dt$. This occurs via the implementation of a cosmic four-vector $\lambda_{\mu}$. The covariant formulation designed in \cite{lazo} has been so far studied in the context of FLRW expansion \cite{Carames:2018atv,Fabris:2017msx} and its applications to Braneworld gravity \cite{Fabris:2018nli} and cosmic strings \cite{Braganca:2018elt}.

In this work we study the aspects of the action-dependent Lagrangian gravitational theory related to the existence and compatibility with astrophysical sources. The field equations of this theory are presented in the section \ref{section2}. Later, in the section \ref{TOV}, we investigate static configurations aiming to describe the interior solution of spherically-symmetric astrophysical compact objects. This analysis imposes strong conditions on the departure from the GR theory represented by the free parameter of the theory contained in a four-vector $\lambda_{\mu}$. Indeed, anticipating the main results of this work, static spherically symmetric objects only exist for a very particular form of $\lambda_{\mu}$. For a reasonable equation 
of state (EoS) for the neutron star interior we plot the equilibrium mass-radius diagrams in order to assess the impact of the available components 
of $\lambda_{\mu}$ into the maximum mass of the neutron star. In the section \ref{coll} the spherical collapse of dust matter in the considered theory is analysed as well as the singularity points which appear. We conclude in the final section.

\section{The field equations of the theory}\label{section2}

The complete set of field equations considered in \cite{lazo} is based on the following total Lagrangian
\begin{equation}
 \mathcal{L}=\mathcal{L}_g+\mathcal{L}_m=(R-\lambda_\mu s^\mu)+\mathcal{L}_m,
\end{equation}
where apart from the Einstein-Hilbert Lagrangian for the geometrical sector one deals with the additional dissipative term $\lambda_\mu s^\mu$ while $\mathcal{L}_m$ is the Lagrangian of
the matter fields. In general, the background four-vector $\lambda^\mu$ depends on the spacetime coordinates, however it can be assumed to be constant. The field $s^\mu$ is an action-density field which 
disappears after the variation of the action such that the modification to the GR counterpart is given by the four-vector $\lambda^\mu$ only. Thus, the field equations 
are given by
\begin{equation}\label{eq}
 G_{\mu\nu}+K_{\mu\nu}-\frac{1}{2}g_{\mu\nu}K=\kappa T_{\mu\nu},
\end{equation}
where we have defined $\kappa=-8\pi G/c^4$, $G_{\mu\nu}=R_{\mu\nu}-\frac{1}{2}g_{\mu\nu}R$ is the Einstein tensor, $T_{\mu\nu}$ is the perfect fluid energy momentum sourcing the field equations 
\begin{equation}
 T_{\mu\nu}=(p+\rho)u_{\mu}u_{\nu}+pg_{\mu\nu},
\end{equation}
where $p$ is the pressure, $\rho$ denotes the energy density, and $u^\mu$ is a normalized $u^\mu u_\nu=-1$ four-vector field (an observer co-moving with the fluid). The symmetric geometric structure $K_{\mu\nu}$ is defined as
\begin{equation}
 K_{\mu\nu}=\lambda_\alpha\Gamma^\alpha_{\mu\nu}-\frac{1}{2}(\lambda_\mu\Gamma^\alpha_{\nu\alpha}+\lambda_\nu\Gamma^\alpha_{\mu\alpha})
\end{equation}
which is constructed from the particular combination of the four-vector $\lambda_{\mu}$ and the Christofell symbols 
\begin{equation}
\Gamma^\alpha_{\mu\nu}= \frac{g^{\alpha\beta}}{2}\left(g_{\beta\mu,\nu}+g_{\beta\nu,\mu}-g_{\mu\nu,\beta}\right).
\end{equation} 
The quantity $K_{\mu\nu}$ (and its trace $K$) represents the geometric structure behind the dissipative nature of such theory. Again, the limit of a vanishing $\lambda_{\mu}$ restores the dissipationless feature of GR. 

\section{Static Spherically-symmetric solution}\label{TOV}
It is worth noting that the modified field equations (\ref{eq}) belong to a class of modified gravity theories whose 
field equations can be written according to the structure \cite{cap1, cap2, cap3}
\begin{equation}\label{mod}
 \sigma(\Psi^i)(G_{\mu\nu}-W_{\mu\nu})=\kappa T_{\mu\nu},
\end{equation}
where $\sigma(\Psi^i)$ represents a parameterized coupling to the gravity being $\Psi$ either a curvature invariant or even other fields which do not 
have a geometric origin, like scalar ones. The symmetric tensor $W_{\mu\nu}$ parameterizes additional geometrical terms which may appear in an specific modified theory under consideration. Clearly, in order to recast Eq. (\ref{eq}) into Eq. (\ref{mod}) one identifies the coupling $\sigma$ as the unity, i.e.,
\begin{equation}
\sigma(\Psi^i)=1,
\end{equation}
while 
\begin{equation}
W_{\mu\nu}=-K_{\mu\nu}+\frac{1}{2}g_{\mu\nu}K.
\end{equation} 

Therefore, since we are interested in static spherically-symmetric solutions, let us consider a spacetime metric of the form
 \begin{equation}\label{ssm}
  ds^2=-c^2B(r)dt^2+A(r)dr^2+r^2d\Omega^2.
 \end{equation}
One may use the procedure for the modified field equations of the form (\ref{mod}) presented in Ref. \cite{aneta} and hence we immediately can write down 
the solutions of the field equations inside a compact object ($r<R_{\star}$, where $R_{\star}$ is a star's radius) such that
\begin{align}\label{sol}
 A(r)&=\left(1-\frac{2GM}{c^2r}\right)^{-1},\\
 B(r)&=e^{\int_0^r \left(\frac{-2 G \lambda_r M \tilde{r}/c^2-4 G M+2 k \tilde{r}^3 p(\tilde{r})+\lambda_r \tilde{r}^2-2 \lambda_\theta \tilde{r}\cot(\theta )}{4 G M \tilde{r}/c^2-2 \tilde{r}^2}\right)d\tilde{r}},
\end{align}
where the mass function is defined as
\begin{equation}\label{mass}
M\equiv M(r)=\int_0^r\left(4\pi r^2\rho -\frac{3\lambda_r \tilde{r}+2\lambda_\theta \text{Cot}(\theta)A}{4 G A}\right)d\tilde{r}.
\end{equation}

We proceed now further by analyzing the modified Einstein field equations (\ref{eq}). It is worth nothing that the traceless part also provides additional information about the geometrical structure of the solutions of the theory. In order to satisfy the traceless part of the modified equation (\ref{eq}), 
\begin{equation}
K_{\mu\nu}=0,\,\mu\neq\nu,
\label{traceless}
\end{equation} 
one obtains 
\begin{equation}
\lambda_t=\lambda_\phi=0.
\end{equation}
This condition reduces the available free parameters of the theory to only two parameters $\lambda_{r}$ and $\lambda_{\theta}$. Also, further analysis of (\ref{traceless}) imposes that such remaining components of $\lambda^\mu$ are related by 
\begin{equation}
 \lambda_\theta\left(\frac{A'}{A}+\frac{B'}{B}\right)+2\lambda_r\cot\theta=0.
\label{AB}
\end{equation}

Let us notice that we have assumed a static spherically-symmetric format for the metric components $A(r)$ and $B(r)$. 
Therefore, the terms including $\cot(\theta)$ should depend on the coordinate $r$ only. Thus, by inspection of the mass function
(\ref{mass}) and (\ref{AB}) it is worth to notice that the existence of radially symmetric mass profiles imposes the functional 
form $\lambda_\theta\sim \cot(\theta)^{-1}$. Applying this requirement to the equation (\ref{AB}) we obtain
\begin{equation}
 \lambda_r=0,\;\;\;\;\;\lambda_\theta=\frac{\lambda_{0}}{\cot\theta},\;\;\;\lambda_{0}=\text{const}.
\end{equation}
Therefore, the dimensionless parameter $\lambda_0$, related to the $\lambda_{\theta}$ component of the four-vector $\lambda_{\mu}$, is the unique parameter of the theory in static spherically symmetric spacetimes.  
Henceforth, the static spherically symmetric solutions of the action-dependent theory are written in a much more simplified form
\begin{align}
 A(r)&=\left(1-\frac{2GM(r)}{c^2r}\right)^{-1},\label{sol2}\\
 B(r)&=e^{\int_0^r\left(\frac{-4 G M(r) /c^2+2 \kappa \tilde{r}^3 p(\tilde{r})-2 \lambda_0  \tilde{r}}{4 G M(r) \tilde{r}/c^2-2 \tilde{r}^2}\right)d\tilde{r}}\label{sol2b}
\end{align}
where the mass function becomes
\begin{equation}\label{mass2}
 M(r)=\int_0^r\left(4\pi r^2\rho -\frac{\lambda_0 c^2}{2 G }\right)d\tilde{r}.
\end{equation}

\section{The interior solution of Relativistic stars in action-dependent Lagrangian theories}
By using the Bianchi identities in the field equations (\ref{eq}), it turns out that the standard conservation law is replaced by
\begin{equation}
 K^{\mu\nu}_{\,\,;\mu}-\frac{1}{2}K^{;\nu}=\kappa T^{\mu\nu}_{\,\,;\mu}.
\end{equation}
Using the above equation and realizing that the theory can be recasted in the form (\ref{mod}) we can apply the results of \cite{aneta} to obtain the modified Tolmann-Oppenheimer-Volkoff (TOV) equation as
\begin{align}\label{m2TOV}
 p'(r)=
 &-\frac{GM}{r^2}(\rho+\frac{p}{c^2})\left(1+\frac{4\pi r^3\left(p-\frac{\lambda_0}{\kappa r^2}\right)}{M c^2}\right)\nonumber\\
 &\times
 \left(1-\frac{2GM}{c^2r}\right)^{-1},
 \end{align}
which together with the mass function (\ref{mass2}) and an appropriate EoS $p=p(\rho)$ will allow to study the equilibrium stellar structure in this model.

We now calculate numerical solutions of the modified version of the TOV equation (\ref{m2TOV}) for a suitable EoS which describes the interior of compact objects like neutron stars.

Neutron stars properties like their masses and radius have been determined and used to constrain the cold dense matter EoS within compact stars. Remarkably, it is worth to mention the discovery of well precisely measured pulsars
with masses close to 2 $M_{\odot}$ as for example PSR J0348+432 \cite{antoniadis}, or even above this value, like PSR J2215+5135 with the mass around $2.27M_\odot$ \cite{huge}. These results lead to the exclusion of soft EoSs which can not reproduce such massive neutron stars.

Although the existence of many distinct models for the NS interior, $2 M_{\odot}$ objects can be reproduced phenomenologically via polytropic EoSs of the form $p = \kappa_0 \rho^{\gamma}$ where $\kappa_0$ and $\gamma$ are fitting constants. Following the toy model presented in \cite{Silbar:2003wm} we adopt an approach in which a pure neutron matter model (i.e., vanishing proton contribution) has been fitted by the quadratic polytropic EoS $\gamma=2$ (in units preserving that $p$ and $\rho$ are given in $MeV/fm^3$) providing therefore $\kappa_0=4.012 \times 10^{-4} fm^3 / MeV$. 

We proceed now by solving numerically the modified TOV system given by the equations (\ref{mass2}) and (\ref{m2TOV}). Each hydrostatic equilibrium configuration is the outcome of a given central density $\rho_0$ which is a necessary input in order to solve this system. By spanning many order of magnitude in the central density, from $10^{-5} - 10^{-1} M_{\odot}/km^3$, one obtains the so called mass-radius diagram.

\begin{figure}[t]
\includegraphics[width=0.4\textwidth]{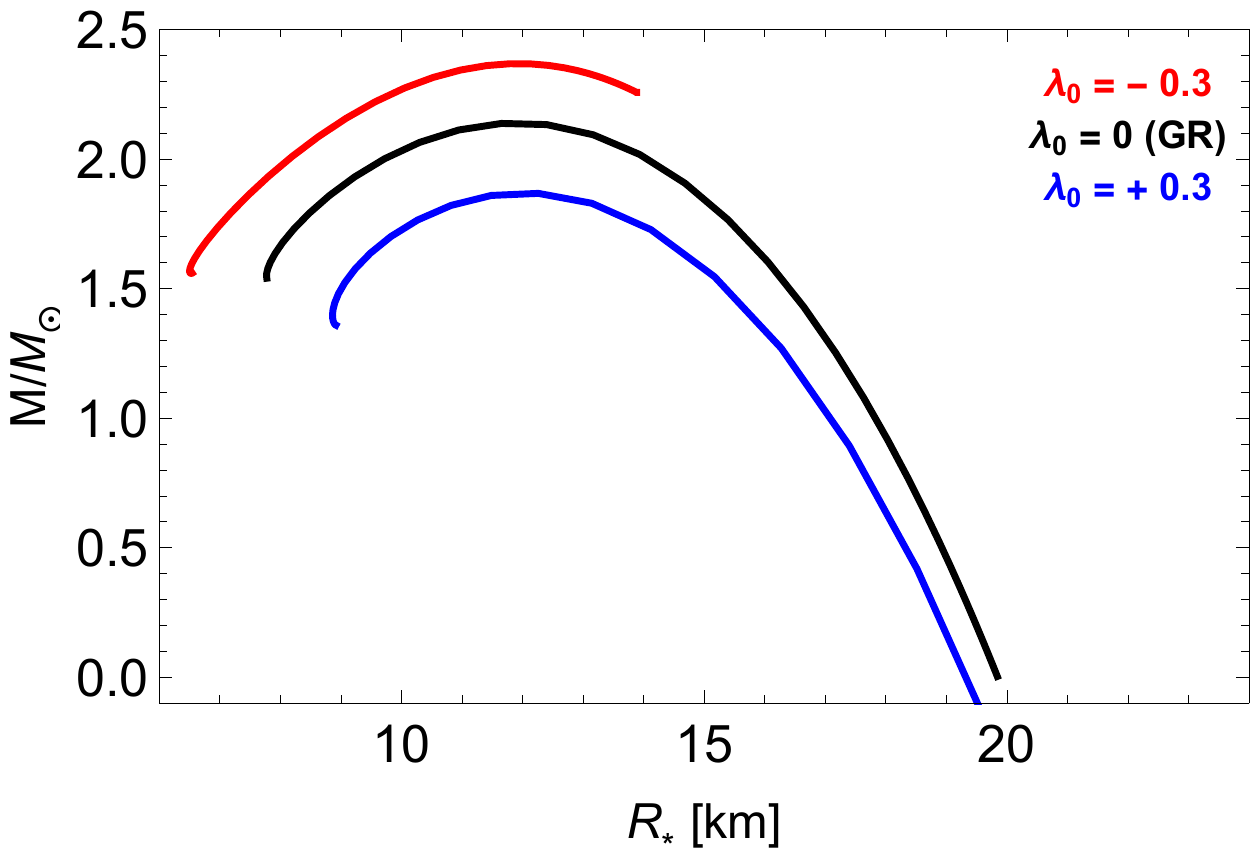}
\caption{Mass-radius diagram. The black line represents the GR configurations. For the red (blue) line the parameter value $\lambda_0=-0.3 (+ 0.3)$ has been adopted.}
\label{fig1}\end{figure}

In Fig.\ref{fig1} we plot the mass-radius diagram for the hydrostatic equilibrium configurations using the above mentioned EoS. Indeed, as seen 
in the black curve, GR predicts $2 M_{\odot}$ objects with radius in the range $\sim 10 - 14 km$. The red line corresponds to the 
value $\lambda_0 = - 0.3$ while the blue line to $\lambda_0 = + 0.3 $. The former simulates the effect of a harder EoS while the latter tends 
to decrease the maximum mass acquired by neuron stars as in the case of soft EoSs. Therefore, positive values of $\lambda_0$ seems to be qualitatively disfavored, 
at least in the framework of the simple polytropic equation we have used. Let us notice that the red curve is not allowed to proceed to high radius values due to numerical instability. As we shall see in the section, the value $\lambda_0=-0.3$ is close to the one for which there is the appearance of a naked singularity.

\begin{figure}[t]
\includegraphics[width=0.4\textwidth]{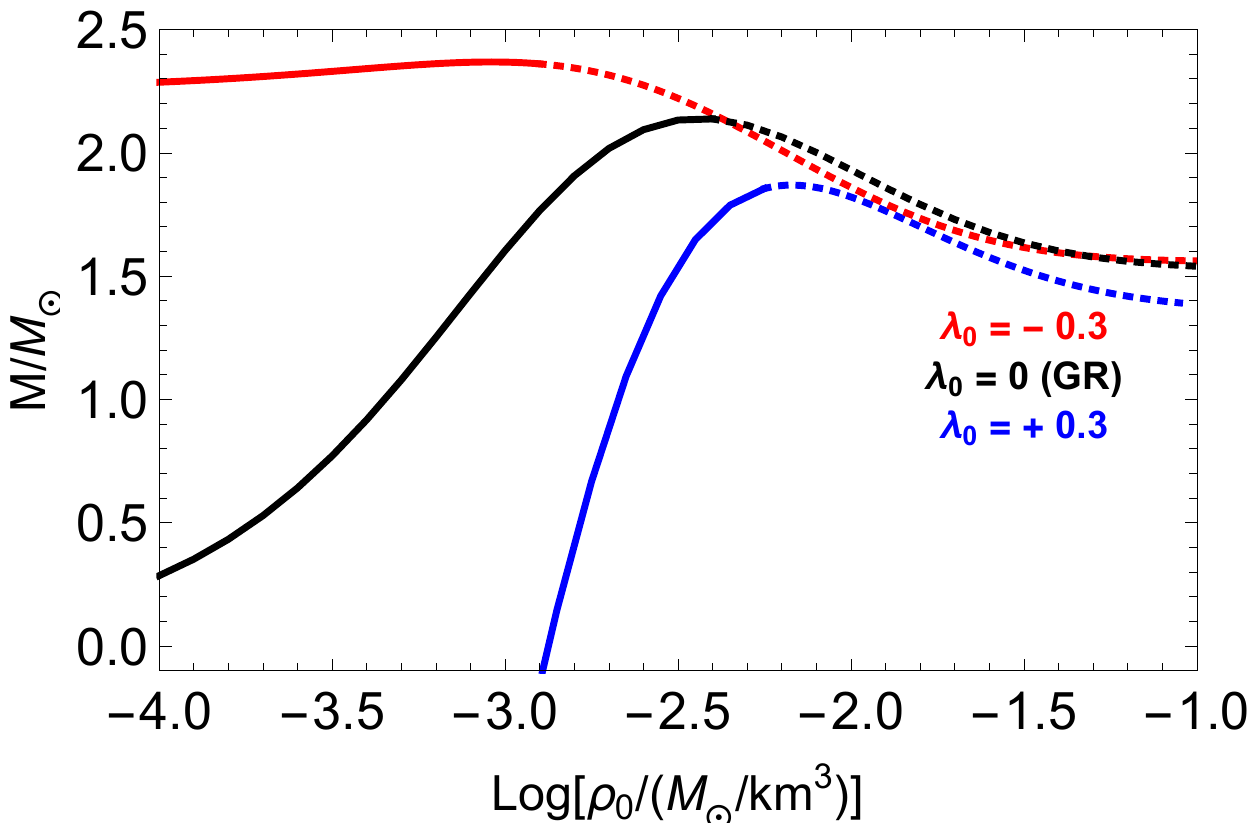}
\caption{Neutron star mass versus its central density. Stable configurations are represented by the solid lines. Unstable configurations shown in the dotted lines. }
\label{fig2}\end{figure}

Fig. \ref{fig2} shows the gravitational mass of the star as a function of its central density. Solid lines of Fig. \ref{fig2} show stable configurations. Beyond the central density at which the maximum mass is obtained the configuration becomes unstable. The latter condition is shown in the dotted lines of Fig. \ref{fig2}.

 \section{Spherical symmetric collapse of dust}\label{coll}
The metric (\ref{ssm}) with the coefficients given by (\ref{sol2}) and (\ref{sol2b}) 
takes the following form in vacuum 
\begin{align}
 ds^2=&-c^2\left(1-\frac{2GM_0+\lambda_0c^2R_\star}{c^2(\lambda_0+1)r}\right)dt^2\nonumber\\
 &+\left(1-\frac{2GM_0+\lambda_0c^2R_\star}{c^2r}+\lambda_0\right)^{-1}dr^2+r^2d\Omega^2\label{metric}
\end{align}
where $M_0=M(R_\star)$. The metric reduces to the well-know GR solution, that is, the Schwarzschild one, when $\lambda_0=0$. 
 
 Before studying the gravitational collapse, it is worth noting that the metric (\ref{metric}) clearly depends on the sign
 of the constant $\lambda_0$. Let us recall that the mass function appearing in this expression is written as
 \begin{equation}\label{masa}
  M(r)=M_0-\frac{\lambda_0c^2(r-R_\star)}{2G}= M_\text{eff}-\frac{\lambda_0c^2}{2G}r,
 \end{equation}
where $M_\text{eff}=M_0+\frac{\lambda_0c^2}{2G}R_\star$. We notice that for $r\rightarrow\infty$ one does not deal with the asymptotically flat 
metric but instead
(if $\lambda_0\neq -1$) the effective metric becomes
\begin{equation}
 d\tilde s^2=
 -c^2dt^2+d\tilde r^2+(1+\lambda_0)\tilde r^2d\Omega^2\label{noflat},
\end{equation}
with $r=\sqrt{1+\lambda_0}\tilde r$.
Moreover, for $\lambda_0<-1$ the metric changes the signature from $(-+++)$ to $(--++)$. We will discuss about that later on in more details.

Allowing $\lambda_0<-1$ would also manifest a pathological feature in the definition of the mass (\ref{masa}): it would grow with 
$r\rightarrow\infty$ because of the last term. Therefore, we find here the analogy to a negative cosmological term
in the anti-de Sitter spacetime which in similar manner introduces negative 
energy density. On the other hand, in the case of the positive $\lambda_0$ we observe a negative effect in the mass function (\ref{masa}) for $r>R_\star$ working 
analogically as a positive cosmological term, that is, a dissipative contribution introducing a repulsive behavior at large distances.

The exterior solution of the spherical symmetric collapse of dust is given by (here $k=\{0,\pm1\}$ is a constant curvature of space)
\begin{equation}\label{metric2}
 ds^2=-c^2d \tau^2+a^2(\tau)\left(\frac{d\sigma^2}{1-k\sigma^2}+\sigma^2d\Omega^2\right),
\end{equation}
which will be matched by using the Darmois junction conditions \cite{darmois} with
\begin{equation}\label{metric3}
 ds^2=-c^2B(r)dt^2 +A(r)dr^2+r^2d\Omega^2,
\end{equation}
with $A(r)$ and $B(r)$ given by (\ref{metric}).
The matching hypersurface $\Sigma$ is given 
by the equation $\sigma=\sigma_0$, with coordinates $(u,\theta,\phi)$ such that the canonical embedding of $\Sigma$ in the outer region is
\begin{equation}
 i_\text{out}:\Sigma\rightarrow \mathcal{M}: (u,\theta,\phi)\mapsto(\tau(u),\sigma_0,\theta,\phi)
\end{equation}
while embedding into the inner region
\begin{equation}
 i_\text{in}:\Sigma\rightarrow \mathcal{M}: (u,\theta,\phi)\mapsto(t(u),r(u),\theta,\phi).
\end{equation}
 Let us notice that $\sigma_0=R_\star$, that is, 
the matching taken on the star's surface is a subclass of possible matchings \cite{fatibene}. Thus, equaling the first and the second fundamental forms of the hypersurface $\Sigma$ obtained in 
two embeddings in the outer (\ref{metric2}) and inner (\ref{metric3}) regions one has the following equations to be satisfied
\begin{align}
 (t_{,u})^2&=-\frac{\xi^2A^2}{B(1-\xi^2A)}(r_{,u})^2,\label{eq1}\\
 (\tau_{,u})^2&=-\frac{A}{(1-\xi^2A)}(r_{,u})^2,\label{eq2}\\
 r^2&=a^2\sigma_0^2,\\
 0&=(r_{,u})^2\frac{A^2\xi^2(AB)_{,r}}{B(1-A\xi^2)^2}\label{eq4},
\end{align}
where $\xi=\sqrt{1-k\sigma^2_0}$. The derivatives with respect to the coordinate $r$ are later on evaluated at $r=r(u)$.

The last equation, since $r_{,u}\neq0$, is satisfied when $AB=d=\text{const}$ which provides the constant $d$ as 
\begin{equation}
 d=\frac{c^2}{1+\lambda_0}.
\end{equation}
Immediately we find from (\ref{eq1}) and (\ref{eq2}) that
\begin{equation}
 (t')^2=\frac{\xi^2A}{B}(\tau')^2
\end{equation}
which is singular for $r_\text{sing}=0$ and
\begin{equation}\label{sin1}
r_\text{horizon}=\frac{2GM_0+\lambda_0c^2R_\star}{c^2(\lambda_0+1)}
\end{equation}
reducing to the Schwarzschild radius for $\lambda_0=0$.

The analysis of the Kretschmann invariant, which is defined $I=R_{\mu\nu\alpha}^{\;\;\;\;\;\;\beta}R^{\mu\nu\alpha}_{\;\;\;\;\;\;\beta}$, shows that the invariant is 
singular only for $r=0$:
\begin{align}
 I&=\frac{1}{c^4 r^4}\Big(r^4 \lambda_0^2 (B'')^2+2 r^4 \lambda_0  (B'')^2+r^4
   (B'')^2\Big.\nonumber\\
   &\Big.+4 r^2 (\lambda_0 +1)^2 (B')^2+8 c^2 (\lambda_0 +1) B\Big.\nonumber\\
   &\Big.+4
   (\lambda_0 +1)^2 B^2+4 c^4\Big),
\end{align}
and thus the singular point (\ref{sin1}) could be an event horizon. We have already excluded at the 
begining of the section the value $\lambda_0\neq-1$ which 
makes (\ref{sin1}) singular.

\section{Conclusions}
We have presented the static spherically-symmetric solution for the action-dependent Lagrangian theories in which the dissipative term behavior is 
ruled by the background four-vector $\lambda^\mu$. It turns out that from the field equations we deal with $\lambda_t=\lambda_\phi=0$ 
while in order to have radially symmetric mass we need to have the radial coordinate $\lambda_r=0$ and $\lambda_\theta=\lambda_0/\cot{\theta}$ where 
$\lambda_0$ is an arbitrary constant. Due to that fact, the interior spherically-symmetric solution has the forms (\ref{sol2}) and (\ref{sol2b}). 
However, for $\lambda_0\neq0$ the vacuum solution (\ref{metric}) is not an asymptotically flat metric (\ref{noflat}) being singular
for $\lambda_0=-1$ and changing the signature for $\lambda_0<-1$ to $(--++)$.

Despite that, for $\lambda_0>-1$ one obtains the modified TOV equation (\ref{m2TOV}) with a modification $\sim-\lambda_0/r^2$ entering 
to the gravitating component of pressure. The numerical analysis of a toy model described by the polytropic equation of 
state $p=\kappa_0\rho^2$ is depicted in the figures \ref{fig1} and \ref{fig2}. Mass-radius diagram \ref{fig1} shows that 
positive values of the parameter $\lambda_0$ decrease the star's parameter while the negative one provides heavier 
stars. The stable configurations for the chosen values of $\lambda_0$ are represented by the solid lines in the 
Fig. \ref{fig2}. Analogically to the GR case ($\lambda_0=0$) the unstable configurations appear beyond the central 
density for which the star's mass reaches the maximum value.

In the section \ref{coll} we discussed the collapse of dust matter in the model. We have used Darmois junction conditions
to match the exterior solution (\ref{metric2}) with our vacuum solution (\ref{metric}). The conditions are realized by the 
set of equations (\ref{eq1})-(\ref{eq4}) which will be satisfied when $AB=\frac{c^2}{1+\lambda_0}$. We are dealing with the 
central singularity point at $r=0$, which is a physical singularity, and an event horizon given by (\ref{sin1}). It is singular 
for the value $\lambda_0=-1$ and how it was discussed, this would be a point in which the metric changes its signature from 
$(-+++)$ to $(--++)$. Furthermore, for $-1<\lambda_0<-\frac{2GM_0}{c^2R_\star}$ one 
would deal with a naked singularity at $r=0$.  Since that problem requires deeper analysis and it is not directly related to stellar objects, which are our main 
interest here, we will leave it for the future work.

\acknowledgments{The authors acknowledge financial support from FAPES, CAPES, CNPq (Brazil).}

\end{document}